\documentclass[lettersize,journal]{IEEEtran}
\usepackage{amsmath,amsfonts}

\usepackage[ruled,vlined]{algorithm2e}

\usepackage{dsfont}

\usepackage{array}
\usepackage[caption=false,font=normalsize,labelfont=sf,textfont=sf]{subfig}
\usepackage{textcomp}
\usepackage{stfloats}
\usepackage{url}
\usepackage{verbatim}
\usepackage{graphicx}
\usepackage{cite}

\begin{document}


\title{Collective Adaptation in Multi-Agent Systems: How Predator Confusion Shapes Swarm-Like Behaviors
}


\author{
    \IEEEauthorblockN{Georgi Ivanov\IEEEauthorrefmark{1}, George Palamas\IEEEauthorrefmark{1}} \\\
    \IEEEauthorblockA{\IEEEauthorrefmark{1}Aalborg University Copenhagen
    \\\ gpa@create.aau.dk}
}


\markboth{Journal of \LaTeX\ Class Files,~Vol.~14, No.~8, August~2021}%
{Shell \MakeLowercase{\textit{et al.}}: A Sample Article Using IEEEtran.cls for IEEE Journals}


\maketitle

\begin{abstract}

Popular hypotheses about the origins of collective adaptation are related to two basic behaviours: protection from predators and a combined search for food resources. Among the anti-predator explanations, the predator confusion hypothesis suggests that groups of individuals moving in a swarm aim to overwhelm the predator while the dilution of risk hypothesis suggests that the probability of a single prey being targeted by a predator is lower in larger groups. In this paper, we explore how emergent behaviors arise from a predator-driven process as an adaptive response to external stimuli perceived as threatening. Moreover, we suggest a predator confusion process to provide a selective pressure for the prey to evolve group formations. We analyze the foraging and prey-predator dynamics evolved in terms of group density and formation, behavior consistency, predator evasion and success rate, and foraging rate. Two agents' perceptual models are compared. A local observation model, where agents can only see what's in their immediate vicinity, and a global observation model, where agents are able to see the predator at all times. Both models were evolved for predator avoidance, foraging and collision avoidance, using reinforcement learning in a simulated game environment. Our results suggest that the dilution of risk factor is sufficient to evolve group formations, and the predator confusion effect could play an important role in the evolution of collaborative behaviors. Finally, we show how variations in the information exchange of this social order can impact the global, collective behaviors.


\end{abstract}


\section{Introduction}

Self-organization describes the spontaneous, often seemingly purposeful, formation of coherent patterns and emergent behaviors of global order from a set of local interactions. This process enables a collective adaptation where a population of animate or inanimate agents continually interact and adjust their behaviors in order to achieve a common goal \cite{bedau2008emergence}\cite{cilliers1999complexity}. Self-organizing systems can be as diverse as crystal growth, swarm behavior in insects and birds or as complex as cooperative strategies in animals \cite{ashby1968principles} and cultural evolution of language in human societies \cite{de2011self}.

Self-organisation has been an important concept within a number of disciplines including cybernetics and AI, opening up exciting new opportunities to work towards an understanding of the complexity of social dynamics in the context of new disciplines such as Artificial Life and Synthetic Ethology. 

Artificial life (AL) draws heavily from self-organizing systems in a variety of contexts. Turing's theory of morphogenesis aims to describe the development of patterns and forms in living systems based on a two-factor reaction-diffusion scheme \cite{turing1990chemical}. Collective behaviors, broadly observed in animal group formations, such as insect swarm and bird flocks demonstrate properties of self-organization \cite{camazine2003self}. In machine learning the self organizing feature map (SOM), proposed by Teuvo Kohonen \cite{kohonen1990self} can generate spatial formations of abstracted internal representations, similar to the topological properties of brain connectivity. 

In this extension of Artificial Life lies the notion of Synthetic Ethology and the evolution of cooperative communication. Synthetic ethology aims at describing the social interactions of multi-agent systems, usually through simulations in an artificial environment, by means of competitive and cooperative strategies \cite{maclennan1993synthetic}. These simulations provide simplicity and control, as well as practical and empirical validation of complex behaviors that emerge from the interaction of these agents with their environment.

Simulating the behavioral patterns of a multi-agent system is the process of recreating the dynamics of large group of entities, unified by a common purpose, environmental conditions or other external stimulus \cite{bedau2008emergence}\cite{zhou2010crowd}.
In this context, self-organization can be observed in the following aspects: i) an agent's autonomy based solely on the agent's behavioral rules, ii) a total increase in order. One of the earliest attempts at engineering an emergent group behavior is the Boid Algorithm proposed by Craig Reynolds. The algorithm is based on the imitation of the flocking behavior of birds and utilizes a set of simple rules, applied to each boid in the flock \cite{reynolds1987flocks}\cite{reynolds1999steering}. 

An increasing interest has surfaced in analysing and visualizing the emergent behavior of large ensembles of agents, simulating crowd dynamics, modeling physical systems, and modeling the behavior of animals and insects, among others. Dewi et al. \cite{dewi2011simulating} propose the use of flocking boids to simulate realistic crowds, Chiang et al. \cite{chiang2009emergent}, demonstrate the use of simulated emergent phenomena to represents virtual crowd behaviors.  

These works typically rely on a complex system of rules intended to guide the virtual creatures towards a very specific behavior \cite{kwasnicka2011flocking}. While the general predictability of this approach is often beneficial when examining specific desired characteristics of group dynamics, it can raise many questions regarding whether these specific choices of environmental conditions contribute optimally to the emergent process \cite{chang2019investigating}. It is possible that this could result in deterministic behavior, leaving little room for discussion with respect to self-organization, and the potential for its nature-like complexity \cite{chang2019investigating}. 

In this study, we examine the behavior of a multi-agent system in a cooperative environment in the context of anti-predatory and foraging tasks. In addition, a confusion predator model is introduced to simulate potential prey-predator interactions. In this model, the presence of other birds nearby can disrupt the process of targeting prey and thus help explain how flocks of birds successfully evade attacks by diving predators \cite{olson2013predator}. We use a multi-agent simulation with a reinforcement learning algorithm to train the agents to perform their roles and then analyze the properties of the resulting multi-agent system. Last, we compare and contrast the results with the Reynolds model in order to examine the influence of predator confusion on agent survival and social behavior.

\section{Background}

\subsection{Emergent behaviors as Survival Tactics}
Emergent behaviors result from composite systems, which exhibit higher level complexity than their sub-parts \cite{macaulay2016riot}. There is strong evidence that formations of flocks of birds or schools of fish, evolved as defense mechanism against predators \cite{morrow1948schooling}. Grouped together they seek to appear as one large entity, confusing the predator and achieving a significant reduction of hunt-down prey \cite{caraco1980avian}. Swarming ants would form colonies which are vastly more beneficial as a survival strategy than rejecting cooperation \cite{sullivan1981insect}. Emergent behaviors mainly occur in non-linear systems, where the output does not match the input in proportion \cite{macaulay2016riot}. Although often perceived as random structures, they are actually governed by deterministic laws. Micro-changes in input data, which would normally be treated as rounding errors, can produce extreme offsets in the final result \cite{macaulay2016riot}. It is worth noting that one property of emergence is entropy, which, when provided with enough time, leads to self-organization through a process known as spontaneous order, which is closely related to the initial conditions of the system \cite{macaulay2016riot}\cite{camazine2003self}. A few of such systems are random graphs, cellular automata, evolutionary algorithms, multiple-agent systems, swarm intelligence, etc. \cite{mamei2006case}.

\subsection{Dynamics of Prey-Predator Models}

\begin{figure}[h]
    \begin{center}
        \includegraphics[scale=.4]{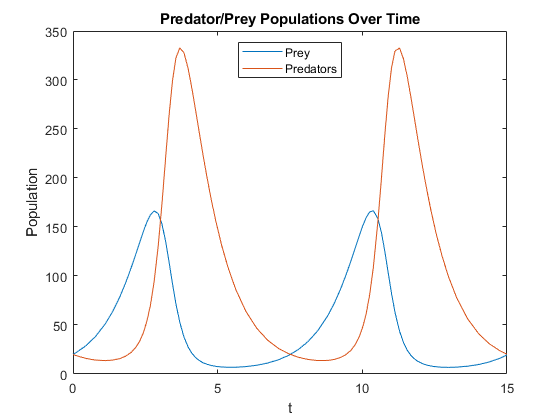}
        \caption{The graph shows the populations of prey and predator based on the Lotka-Volterra equations. The graph shows the cyclic relationship between the populations and their hysteresis, or delayed dependence between them.}
        \label{fig:prey-predator}
    \end{center}
\end{figure}

The purpose of this section is to review various prey-predator hypotheses, often used to analyze interactions between populations within ecosystems, specifically to explore mutualism and how predator behavior may influence the evolution of prey groups.
Lotka-Volterra equations, also known as the predator-prey equations, represent a pair of nonlinear differential equations evolving through time used to describe the dynamics of biological systems where two species interact, one as a predator and the other as a prey \cite{lotka2002contribution}. 
The Lotka–Volterra model can be used to predict the dynamics of competitive behavior between two species, or between a predator and a prey species, and to determine equilibrium population sizes (see figure \ref{fig:prey-predator}). 

However, the model makes several assumptions. The prey population always has enough food, the predators have an unlimited appetite, and the environment does not change to benefit the species. Adaptation of prey systems by predators exhibits similar ecological characteristics as a reaction-diffusion process \cite{garvie2010spatiotemporal}. Typically, the dynamics of these systems, including their relationships, can be described using a pair of coupled reaction–diffusion equations. Other systems exhibiting similar spatio-temporal dynamics are spiral waves \cite{winfree2001geometry}, target patterns \cite{kopell1981target} and localized solitary pulses (oscillons) \cite{vanag2007localized}. Currently, the study of such systems is very active and there are many unanswered questions regarding their generalizability \cite{sherratt1997oscillations}. 

\subsection{Anti-predator Adaptation Behaviors}

There are two popular hypotheses about collective adaptation's origin: A joint search for food resources \cite{witkowski2016emergence}\cite{haley2014exploring}\cite{torney2011signalling} and cooperation in predator avoidance \cite{oboshi2003simulation}\cite{olson2013predator}\cite{haley2014exploring}. According to the first hypothesis, individuals help individuals with whom they are related. The second hypothesis is based on the notion of mutualism, which holds that interactions between two species are mutually beneficial. Neither of these hypotheses are mutually exclusive and both may contribute to the evolution of cooperative behavior. 
Further evidence of this can be found in the fact that collective behavior is common in taxa that are preyed upon, such as insects, fish and birds. Other less popular hypotheses assert that grouping strategies could enhance mating efficiency or even energy efficiency, by improving aerodynamics or hydrodynamics \cite{demvsar2016balanced}.  

\subsubsection{The predator confusion hypothesis} 
This hypothesis proposes that multiple individuals moving in a swarm attempt to overwhelm predators \cite{kunz2006prey}\cite{olson2013predator}. An accumulation of visually similar prey that display similar behaviors may confuse the predator, causing it to lose track and waste time and energy. Furthermore, prey that seems to deviate from the norm, such as sick or elderly individuals, may be picked out by predators more frequently.
\subsubsection{The selfish herd hypothesis} 

This suggests that individuals form groups in an attempt to increase their chances of survival \cite{mills1982satiation}. Hence, aggregations are formed by selfish individuals who attempt to place their conspecifics between themselves and the predator. This results in a perpetual motion towards the center of the group. \cite{olson2016evolution}. 

\subsubsection{The many eyes hypothesis} In the Many Eyes Hypothesis, more eyes provide more efficient scanning of the environment and can detect predators early, while at the same time reducing the time it takes a person to scan the environment \cite{haley2014exploring}\cite{ruxton2008application}.

\subsubsection{The dilution of risk hypothesis} This implies that a predator's probability of targeting a single prey is lower in large groups and that larger groups should be less vulnerable to predators since their attention is diverted away from an individual prey to a larger area of prey. \cite{demvsar2016balanced}.
Despite the dilution of risk factor being sufficient to facilitate group formation, predator confusion effect may play a larger role in the evolution of swarm behavior \cite{olson2016evolution}\cite{demvsar2016balanced}.

Moreover, prey-predator co-adaptation \cite{olson2013predator} may not only provide a selective pressure for the prey to evolve a swarm behavior, but it could also provide an evolutionary pressure for the predator to adapt in a way that minimizes the confusion effect, such as narrowing its angle of view to become less sensitive.
However, in this paper we only deal with prey-driven predator models because, in general, these are the simplest forms of predator–prey dynamics. This is because in such models the prey population is assumed to be in excess of what the predators can consume, so that when a predator captures a prey there is a corresponding number of new prey available to replace the ones consumed. This means that the population numbers of both predator and prey remain constant in time. Such models are known as closed models. In contrast, when the prey population is less than the predators can consume, the predator population will decline in time. Although it is possible to consider such co-self-organizing systems, they also tend to settle into "mediocre stable states" where neither predator nor prey performs well, nor does it improve \cite{ficici1998challenges}.

\subsection{Simulating Multi-agent Systems}

Simulating Multi-Agent Systems (MAS) is important because it facilitates a better understanding of a variety of phenomena such as predator-prey interactions and their dynamics.

In MAS, agents are autonomous or semi-autonomous, display dedicated behavior, and interact with each other and their environments. Based on the application, the interaction among agents may be described in different ways, but a general approach is to assume that each agent has a set of sensors that are used to detect the presence or absence of an event or action, or any other information about the environment.

In general, these agents exhibit universal behavior that can be observed at a macro level. In contrast to equation-based modeling techniques such as Lotka-Voltera, MAS simulations offer an intriguing alternative to represent and model real-world or virtual systems which may be broken down into individuals that interact with one another \cite{parunak1997go}.

MAS is principally based on defining a set of rules or instructions to be followed by the agents. The rules are usually defined by the designer of the MAS, who is familiar with the physical-mathematical model of the system. However, this approach is not appropriate when trying to understand highly complex ecological phenomena, since it is difficult to define these rules a priori.

In order to resolve this problem, one method is to use a global optimizer, such as a genetic algorithm, where the agents learn the rules through an evolutionary process by interacting with their environment \cite{olson2013predator}. In order to evolve and adapt an agent to its environment, an evolutionary approach is based on defining a fitness function to evaluate its performance. As evolutionary systems are global in nature, they can suffer from a premature convergence to a solution or from a lack of stability due to an inability to formulate a global strategy. Furthermore, the fitness function should be carefully defined so as not to bias the evolutionary process in any particular direction.

Reinforcement learning (RL) may represent an alternative approach to an evolutionary approach. RL offers a better balance between exploitation and exploration and is therefore most effective for behavior modeling in highly interactive environments, where the performance of the system is strongly affected by its context. For agents to be sufficiently motivated, reward system and observations should be carefully crafted because they are the key to successful training of any RL agent \cite{schulman2017proximal}. Moreover, the agents should be capable of learning an effective policy in order to cope with uncertainty.


Kwasnicka et al. \cite{kwasnicka2011flocking} used a genetic algorithm to evolve the behaviors of a multi-agent system. Their model introduces a single predator and a health control system for the entire population. According to the authors, the introduction of such diversity in the environment increases the system's entropy to the point where no classifiable behavior can be observed. Based on thorough research on the initial rules and conditions that facilitate grouping behavior, it has been concluded that there is no clear explanation of the emergence process and that further research is required to understand the phenomenon \cite{kwasnicka2011flocking}.

Furthermore, Chang et al. \cite{chang2019investigating}, Brak et al. examine the behavior of agents in a fear-contagion environment against a flocking boid control environment. Their evolutionary environment aims to move a herd of sheep using only one method of control, which is represented by a shepherding dog. According to their results, a grouping behavior seems possible, but unsatisfactory when it comes to matching or exceeding the results obtained in the control environment. The authors conclude that further investigation is required to determine the exact causality relationship between the rules of the environment and the produced outcome. 

Moreover, Hahn et al. \cite{hahn2019emergent} show with their 'SELFish' approach (Swarm Emergent Learning Fish) that emergent flocking behaviors can be achieved through an escape-based approach and a simple reward system in a reinforcement learning environment, where the agents' only motivation is to survive as long as possible. The evolution of group behavior rather than selfishly pursuing the opposite direction, which would maximize the reward, is seen as a Nash equilibrium for the given environment \cite{hahn2019emergent}. This can be explained by the Prisoner's dilemma - multiple participants acting in their own self-interest with the intention of achieving the best possible outcome at the expense of the other participant, yet neither party achieves the best outcome for themselves alone \cite{hamburger1973n}. Furthermore, the authors suggest that additional investigation is required and that future developments, such as adding walls or obstacles to the currently infinite terrain, may result in even more complex emergent behavior.


\section{Methods}

Our goal is to examine the behavioral effects of intelligent agents in a cooperative bio-inspired setting, including foraging, predator avoidance and obstacle avoidance. Based on the research conducted to understand what catalyst conditions contribute to the emergence of flock-like phenomena in nature, the underlying role of predator-prey interactions has been identified. A number of previous attempts have been made to investigate this symbiosis, but various facets of its complexity remain unexplored. Adding to these findings, we examine the effects of a static, in terms of adaptation, predator behavior, and the consequently manifested behaviors of their prey.

\begin{figure}[h]
    \begin{center}
        \includegraphics[scale=.249]{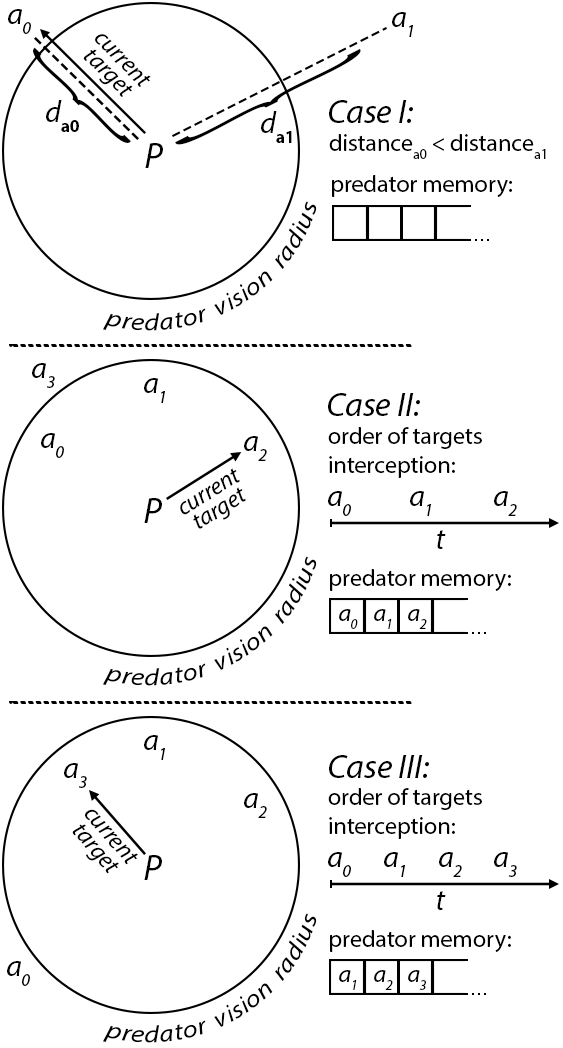} 
        \caption{Flowchart showcasing predator (\textit{P}) behavior in 3 possible cases. In Case I, all agents ($a_0$, $a_1$) are outside the predator's vision radius, thus targeted is the closest one and the predator memory is empty. In Case II, 3 agents ($a_0$, $a_1$, $a_2$) are within its vision radius and one ($a_3$) is outside, thus he is not registered within the predator's memory yet. Since there are more than 1 agents within the predator's vision radius, they are targeted by order of interception. Therefore, it is $a_2$ that is currently being chased. Case III depicts a switch in the predator's targets, as now $a_3$ has entered its radius and $a_0$ exited it. Thus, $a_3$ has been added to the predator's memory and $a_0$ has been removed, respectively. Current target is switched to $a_3$, as he is now the last to enter the predator's vision.}
        \label{fig:flowchart}
    \end{center}
\end{figure}

This paper describes two different models that differ only in their observations - the Global Observation Model (GOM) and the Local Observation Model (LOM). Specifically, the main difference is in how restricted the agents are with regard to the amount of information they can perceive about their surroundings. We compare the results from both implementations to the standard flocking boid behavior proposed by Reynolds (see figure \ref{fig:flowchart}).  The Reynolds flocking algorithm was selected as the control environment because of its behavioral similarity to natural flocking, its elegant simplicity and ease of use \cite{reynolds1987flocks}.

\section{Implementation}

\subsection{Reinforcement learning for intelligent agents}

There are a variety of approaches to the multi-agent simulation problem, including genetic algorithms, particle swarm optimization, ant colony optimization, and self-propelled particles, among others. \cite{rossi2018review}. Evolutionary algorithms are a well-established approach in this context, since they offer a flexible means of approximating a solution without making any assumptions about the "general fitness landscape" \cite{back1993overview}. Because these algorithms are generally aimed at exploitation, they tend to get stuck on local-minima. The use of reinforcement learning algorithms offer a better balance between exploitation and exploration when dealing with complex dynamic systems \cite{coggan2004exploration}. Sutton et al. proposes a solution to train intelligent agents by using a particular action-state-reward causality \cite{sutton2018reinforcement}. 

A key difference between this approach and others is its \textit{memorylessness}, which is known as the Markov property. The environment operates in predefined time-steps, and the agent finds itself in a different state $s$ at each step and can choose between different actions $A$ at each step. As a response, the environment will transition into a new state $s^{\prime}$, considered random from the perspective of the agent, rewarding him with reward $R_a(s, s^{\prime})$. The new state $s^{\prime}$ is directly affected by the chosen action $A$, which is the result of the previous state $s$. Thus, the decision making process is conditionally independent of past states beyond $s$, satisfying the aforementioned property \cite{sutton2018reinforcement}. The agent determines how desirable a state is through its value function, which considering the action-state dependency, can be defined as:

$$\mathcal{V}_\pi(s) = E[R] = E[\sum_{t=0}^\infty\gamma^tr_t | s_0 = s]$$

where \textit{R} is the expected return at a state \textit{s} following a policy $\pi$, $r_t$ is the reward at time step \textit{t} and $\gamma$ [0,1) is the discount rate defining the correlation between the weight of future and immediate events \cite{sutton2018reinforcement}.

Proximal Policy Optimization (PPO) is a good choice of a policy gradient method because of its ability to adjust policies towards the goals of a user, its ease of use, its good performance, and because it allows for the design of policies that adapt to dynamic environments \cite{schulman2017proximal}. PPO's main advantage over more traditional approaches lies in the size of the policy updates, which can be an order of magnitude larger than traditional stochastic gradient descent methods. In contrast, PPO uses a multitude of smaller, more frequent policy updates, combating the issue of single, too large updates derailing the policy. Due to this, PPO updates are applied to the full policy network, instead of just some localized representation of the policy. In turn, this reduces the amount of noise introduced by the updates, which is important when learning a policy in an unknown environment. The main idea behind PPO can be expressed as follows.

$$L^{CLIP}(\theta) = \hat{\mathds{E}_t}[min(r_t(\theta)\hat A_t, clip(r_t(\theta), 1 - \epsilon, 1 + \epsilon)\hat A_t)]$$

where $L^{CLIP}(\theta)$ is a conservative \textit{lower bound} iteration policy, $\mathds{E}_t[...]$ is the expectation averaging over the batch of samples used for the update, $r_t(\theta)$ is a probability ratio, $\hat A_t$ is an advantage function estimator (instead of a Q function) at a time step \textit{t} and $\epsilon$ [0.1, 0.3] is a hyper-parameter, influencing how rapidly the policy can evolve during training \cite{schulman2017proximal}. In more detail, it denotes the threshold between old and new policies during gradient descent updates and allows manual adjustment of the algorithm's stability \cite{juliani2018unity}. The lower bound policy iteration refers to a \textit{pessimistic bound} or one that finds the minimum of the objective expressed above. This allows to distinguish between probability ratios that improve the objective and those that make it worse \cite{schulman2017proximal}.  

\subsection{Environment}

The MAS environment hosts the prey, the predator and the food sources. The only learner in the system is the prey, which is associated with a brain, a neural network that controls its behavior. A policy is formulated based on how the agent interacts with the environment. For the best results, the hyper-parameters of the neural network have been tuned by hand (see table \ref{tab:hyperparam}). 

The environment's active space is defined by a rectangular floor surrounded by walls. The food pieces are placed at random positions, and replaced as soon as they are consumed. In choosing the habitat size, it was observed that either too large or too small an area leads to either overpopulation or inability to utilize the entire usable field effectively, thereby resulting in behavioral artefacts.

\begin{table}[!t]
\centering
\caption{Both models were fitted with the following \\hyperparameters
}
\label{tab:hyperparam}
\scalebox{0.9}{
\begin{tabular}{|c|c|}
\hline
\begin{tabular}[c]{@{}l@{}}trainer: ppo\\     batch\_size: 128\\     beta: 5.0e-3\\     buffer\_size: 2048\\     epsilon: 0.2\\     hidden\_units: 128\\     lambd: 0.95\\     learning\_rate: 3.0e-4\\     learning\_rate\_schedule: linear\\     max\_steps: 5.0e8\\     memory\_size: 256\end{tabular} & \begin{tabular}[c]{@{}l@{}}normalize: false     \\     num\_epoch: 3\\     num\_layers: 2\\     time\_horizon: 64\\     sequence\_length: 64\\     summary\_freq: 1000\\     use\_recurrent: false\\     vis\_encode\_type: simple\\     reward\_signals:\\         \phantom{M}extrinsic:\\             \phantom{MM}strength: 1.0\\             \phantom{MM}gamma: 0.99\end{tabular} \\ \hline
\end{tabular}}
\end{table}

\subsection{Reward system and observations}

\begin{figure}[h]
    \begin{center}
        \includegraphics[scale=.2]{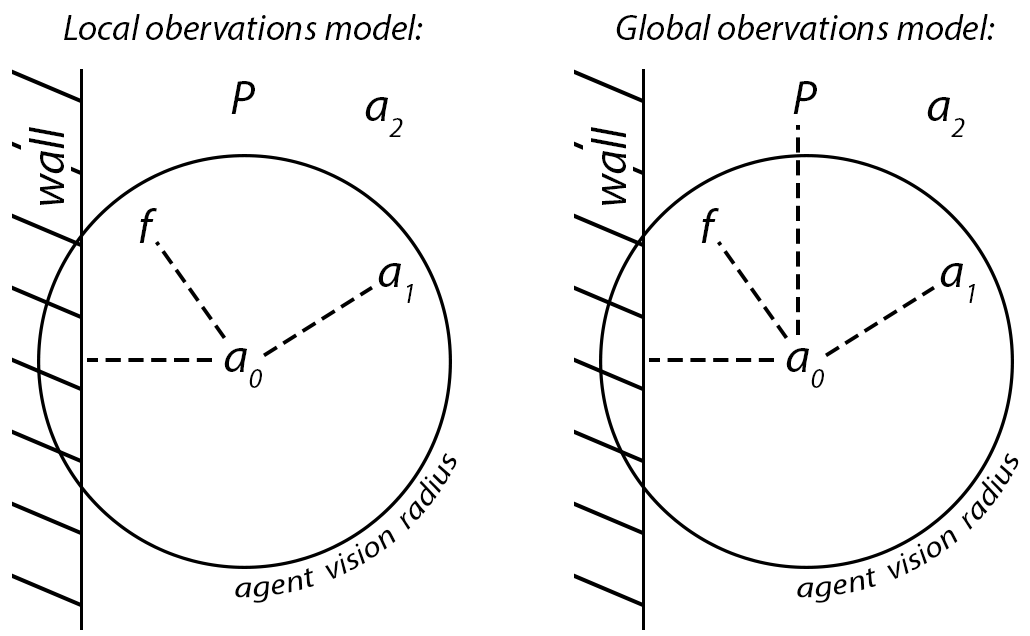} 
        \caption{Difference in the observations available for the two models. Both can perceive walls, food (\textit{f}) and other agents ($a_1$, $a_2$, ..., $a_n$), as long as those are within their observation radii. The global observations model can additionally perceive the predator at all times.}
        \label{fig:observations_graph}
    \end{center}
\end{figure}

\begin{figure}[h]
    \begin{center}
        \includegraphics[scale=.75]{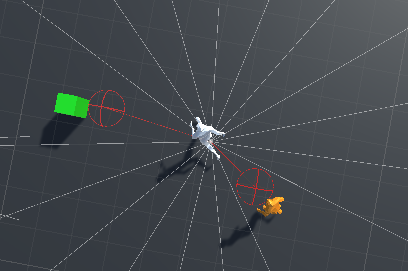} 
        \caption{Agent (white) ray casting observations for food (green) and predator (orange).}
        \label{fig:raycastHit}
    \end{center}
\end{figure}

Each flock member reacts only to events occurring within a certain predefined radius, ignoring everything outside his narrow scope of perception. The LOM model follows this principle, but the GOM agents can see 4 additional metrics related to the predator, at all times. Agents perceive the environment through a nine feature vector consisting of the: agent's position, velocity (x,z) and forward vector; predator's position, velocity (x,z) and forward vector and the distance between the two (see table \ref{tab:rewardSystems}). Using ray casting, the agents can detect four different objects - other agents, predators, food, and walls
(see figures \ref{fig:observations_graph} and \ref{fig:raycastHit}). As a result, the agents have no blind spots. In this regard, the rays represent a circular vision radius around each agent, similar to that used by Reynolds to describe the agents' perception of their surroundings. 

In their respected work, Hahn et al. consider agents as circular objects, that can turn clockwise or counterclockwise to change their forward facing direction \cite{hahn2019emergent}. Due to lack of physical bodies, the agents can move freely, even through each other. We extend this approach to use solid bipedal humanoid bodies for the agents. In consequence, each agent has to manage close proximity interactions with its flock mates in order to avoid collisions. Collisions can restrict prey's movement, making them more vulnerable to predators' attacks. In addition, instead of being able to change directions instantly, the agents need to take several steps to make a turn, requiring a more complex control input for quick and efficient navigation. 

\begin{table}[!t]
\caption{Reward Functions}
\centering
\label{tab:rewards}
\begin{tabular}{|c||c|}
\hline
Foraging & 0.5 \\ 
\hline
Caught by predator & -1 \\
\hline
Wall collision & -0.5 \\ 
\hline
\end{tabular}
\end{table}

\begin{table}[!t]
\caption{Observation spaces used by each MAS model}
\centering
\label{tab:rewardSystems}
\begin{tabular}{ccccc}
\cline{1-3}
\multicolumn{1}{|l||}{LOM} & \multicolumn{1}{l||}{\begin{tabular}[c]{@{}l@{}}- agent position\\ - agent velocity\\ - agent orientation\end{tabular}} & \multicolumn{1}{l|}{\begin{tabular}[c]{@{}l@{}}18 rays with a \\ total spread\\ of 170 degrees\end{tabular}}\\  \cline{1-3}
\multicolumn{1}{|l||}{GOM} & \multicolumn{1}{l||}{\begin{tabular}[c]{@{}l@{}}- agent position\\ - agent velocity\\ - agent orientation\\ - predator position\\ - predator velocity\\ - predator distance\\ - predator orientation\end{tabular}} & \multicolumn{1}{l|}{\begin{tabular}[c]{@{}l@{}}18 rays with a \\ total spread\\ of 170 degrees\end{tabular}} \\  \cline{1-3}
\end{tabular}
\end{table}

Agents receive a reward of 0.5 for feeding, -1 for being caught by a predator, and -0.5 for hitting a wall (see table \ref{tab:rewards}).

\begin{figure}[h]
    \begin{center}
        \includegraphics[scale=.2]{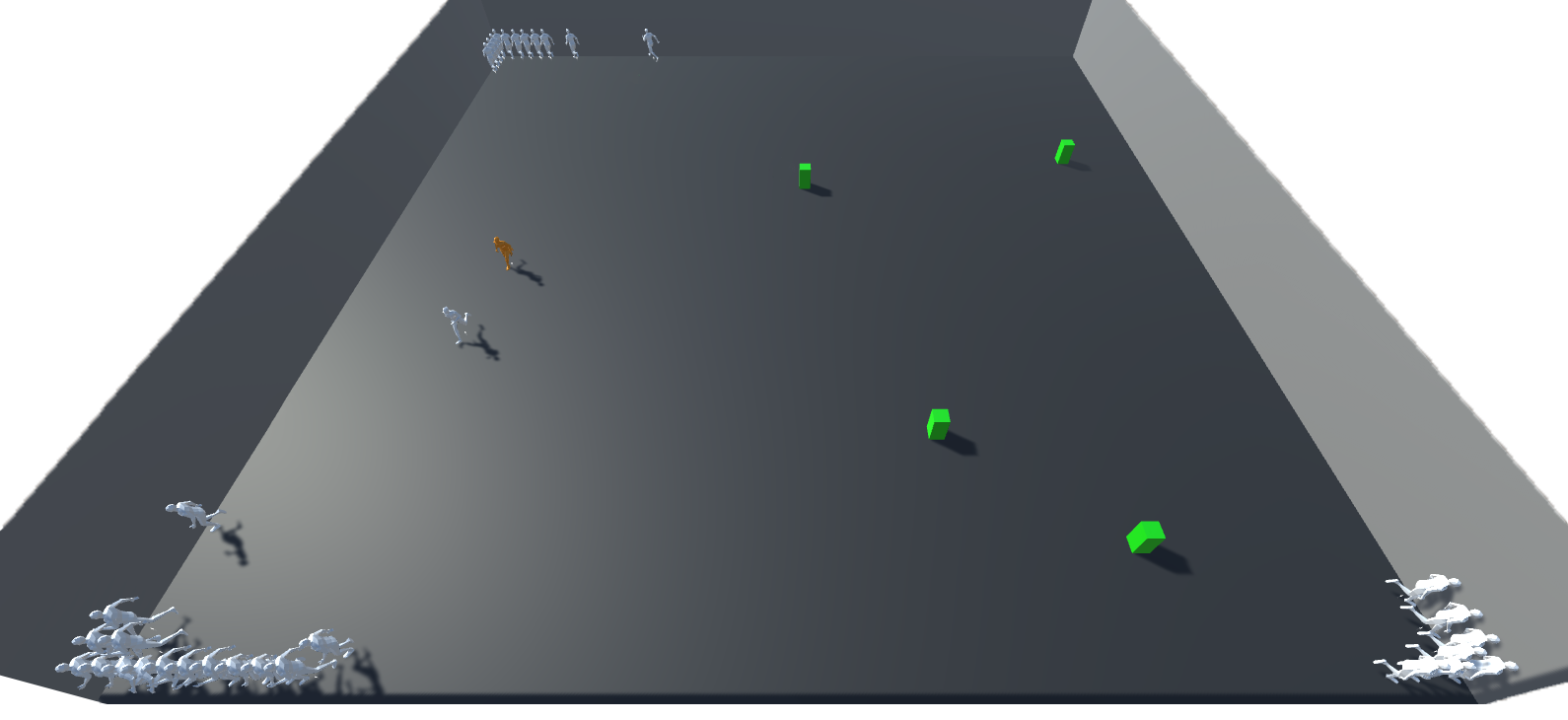} 
        \caption{Lazy behavior of 'picking a corner' adopted by the agents (white), due to absence of a wall-collision penalty.}
        \label{fig:cornerpicking}
    \end{center}
\end{figure}

These values were chosen to balance the evolutionary pressure from different parts of the environment, and they were normalized in the range [-1,1]. Since a predator is the main catalyst behind the agents' behavior, the negative reward for being caught by it should be twice as large as the one for colliding with a wall. The size of the reward for feeding in this case is essentially irrelevant, as long as it is above 0, since this is the only way for the agents to obtain a positive reward. The agents are penalized for hitting the walls, which defines the boundaries of the exploitable area. In the absence of such a penalty, they regress to a lazy behavior of 'picking a corner' until the end of the learning episode, which does not result in any significant progress in policy updates. (see figure \ref{fig:cornerpicking}). It is also equivalent to the method that is more commonly used when dealing with dynamic environments, which rewards idle virtual entities with small negative rewards.\cite{hahn2019emergent}. Both approaches result in motivated behavior of the agents that resembles hunger. From the perspective of the RL model, which can get both positive and negative rewards, getting a neutral reward, which is equal to 0, is the next most desired state after receiving positive rewards. However, since reaching a state with positive rewards entails a higher effort cost, the agent will always opt for the more easily attainable neutral reward \cite{coggan2004exploration}. As currently configured, this means that agents do not move until just before the predator approaches, leading to no learning progression. A penalty such as this ensures that the environment's dynamic equilibrium is maintained.

\subsection{The Predator Confusion Process }

A predator will only chase the closest prey if there are no other prey within its vision radius. If there are more than one, it will go after the last that entered the vision radius. Prey can escape by moving closer to other flock members and thus 'confusing' their attacker. 
Through this process, the prey exploits the predator's inability to process sensory information over an extended period of time by overloading it with sensory stimuli (see pseudo-code in \ref{predPseudoCode}). Hahn, et al. \cite{hahn2019emergent} provide predators with a 50\% chance of targeting each new prey that enters their vision radius in order to create a cognitive overload effect, allowing the agent that was previously being chased to flee. The agent will learn that it is more beneficial to stay close to the flock and not allow itself to be isolated, thereby resulting in tight formations and grouping. The disadvantage of randomizing the predator's choices is that introduces inconsistencies into their behaviors and does not offer long-term solutions.

\subsubsection{Predator's Memory Model}
To avoid the problem stated above, we implement a short-term memory model for the predators, allowing them to remember the prey that entered their vision radius last. The predator will now target the last prey that entered its vision radius, thus creating a consistent behavior that looks more realistic. When the current prey exits the vision radius then the predator will go after the last prey that entered his vision radius. That will allow more direct, linear and predictable chases.

The memory of the predator gets expanded each time a new prey enters its vision range, and each time a prey leaves or gets eaten, it is removed from it. Therefore, the larger an agent's memory is, the more often they escape being eaten because the predator got confused and attacked another flock member during the chase. As a result of the predator's behavior, prey evolves a behavior that makes them seek 'safety in numbers', thus becoming a catalyst for collective action. The agents have been implemented with a maximum velocity 20 per cent lower than the predator's, preventing them from regressing into a behavior of constant circular chasing.



\begin{algorithm}
\SetAlgoLined
 Initialize global memory list\;
 Initialize local memory list\;
 \ForEach{prey}{
 \eIf{Distance between predator and prey \textless= predator vision radius}{
 Add prey to local memory\;
 \uIf{global memory !contains prey}{
 Add prey to global memory\;
 }
 }{
 \uIf{global memory contains prey}{
 Remove prey from global memory\;
 }
 }
 \eIf{local memory size \textless 2}{
 Clear global memory\;
 Return direction towards closest prey in the entire field\;
 }{
 \eIf{global memory size \textgreater 0}{
 Return direction toward prey that has last entered predator vision radius\;
 }{
 Return no changes in direction\;
 }
 }
 }
 \caption{Predator functionality}
 \label{predPseudoCode}
\end{algorithm}

\subsection{Flocking boid control environment}

A particular number of neighbours will be taken into account at any given time by each agent, depending upon the size of its vision range. A cohesion vector is calculated by averaging their positions to create one point of interest, ensuring the agent remains close to its neighbors. During alignment, the average of the forward facing vectors of the neighbours is taken and the agent's orientation is adjusted accordingly. This guarantees similar orientation across all agents. In order to avoid collisions, a second, typically smaller radius is defined. In order to avoid collisions, the agent monitors how many neighbors get too close. The average position of those is calculated and the agent is redirected away from that location.
The cohesion (\ref{eqn:coheq}), alignment (\ref{eqn:aleq}) and avoidance (\ref{eqn:aveq}) behaviors' logic is computed with the following set of equations, where $C$, $Align$ and $Avoid$ are the respective composite vectors for the current agent $A_c$, $N$ is the total number of neighbouring agents $A$ of that agent and $A.pos$ and $A.forward$ are the position and local forward vector of each neighbour:

\begin{equation}
\label{eqn:coheq}
    C = \frac{\Sigma_{i=1}^{N}A_i .pos}{N}
\end{equation}
\begin{equation}
\label{eqn:aleq}
    Align = \frac{\Sigma_{i=1}^{N}A_i .forward}{N}
\end{equation}
\begin{equation}
\label{eqn:aveq}
    Avoid = \frac{\Sigma_{i=1}^{N}A_c.pos - A_i .pos}{N}
\end{equation}

The implementation was based on the ©Unity game engine and the Ml-agents library \cite{juliani2018unity}.

\section{Results}

\begin{figure}[htb]
\centering
  \includegraphics[scale=.35]{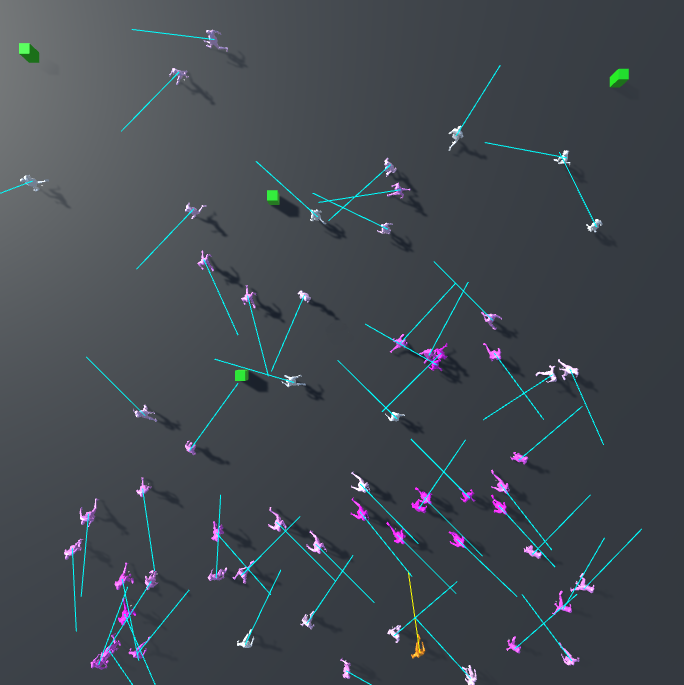}
\\[3px]
  \includegraphics[scale=.35]{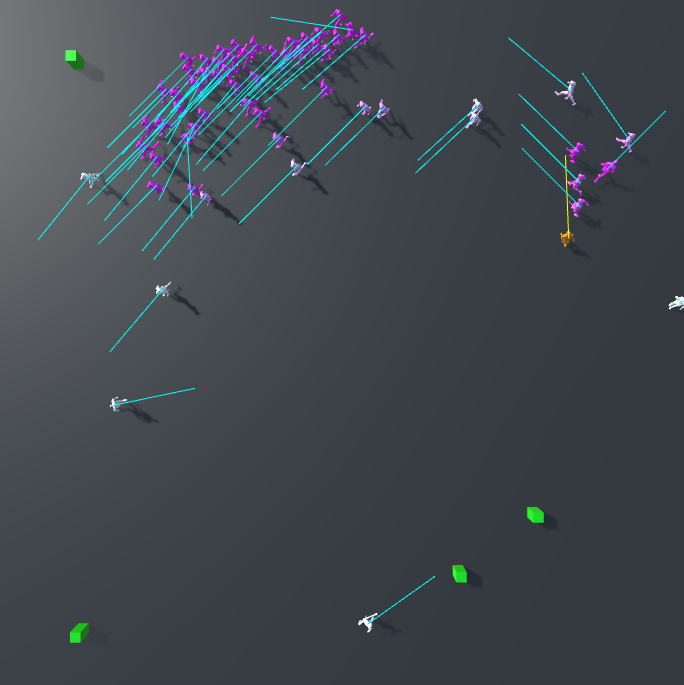}
 \\[3px] 
  \includegraphics[scale=.35]{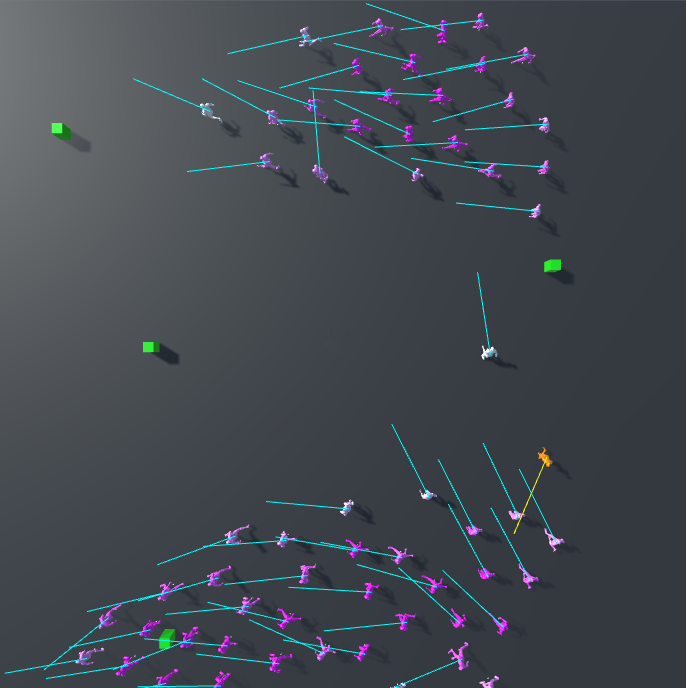}
\caption{An overview of the performance of all three environments. Predator is highlighted in orange. The color interpolation is used to mark crowd members according to the size of their local group - white for 0 neighbours, or magenta when an agent is completely surrounded. Cyan represents forward vectors of the agents. Green boxes represents food.}
\label{fig:examplesBehavsEnv}
\end{figure}

A comparison was made between the manifested behavior of the LOM and GOM models versus a control environment (see figure \ref{fig:examplesBehavsEnv}). Both models differ only in the number of observations they receive for the environment. Reynolds' boid algorithm is used as the control environment. A table containing the collected metrics can be found at \ref{tab:colMetr}.

\begin{table}[h]
\centering
\caption{Metrics monitored during the experimental procedure
}
\label{tab:colMetr}
\scalebox{0.99}{
\begin{tabular}{|l|}
\hline
\begin{tabular}[c]{@{}l@{}}Distance from the center of weight of the nearest group\end{tabular} \\ \hline
Distance from the predator \\ \hline
Number of neighbors for any given agent \\ \hline
Catch rate of the predator  \\ \hline
Predator's memory size \\ \hline
\begin{tabular}[c]{@{}l@{}}Angular errors between ideal and observed behaviors.\end{tabular} \\ \hline
\end{tabular}}
\end{table}

\begin{figure}[h]
\centering
    \includegraphics[scale=.25]{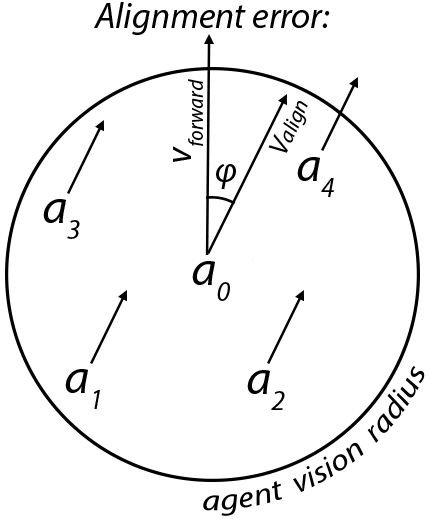}
    \includegraphics[scale=.25]{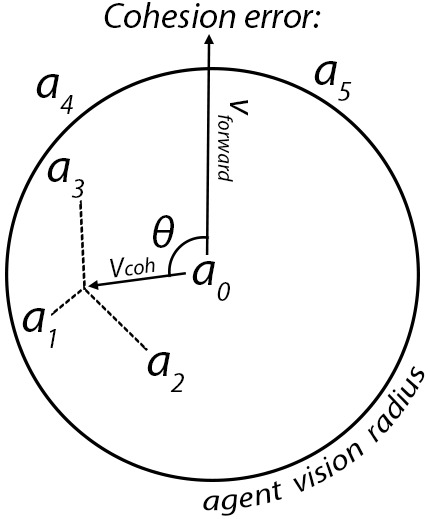}
    \includegraphics[scale=.25]{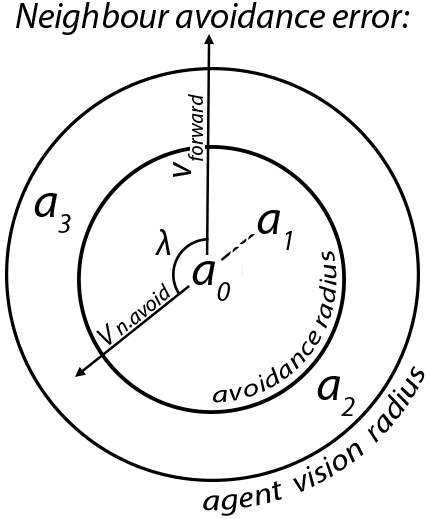}
    \includegraphics[scale=.25]{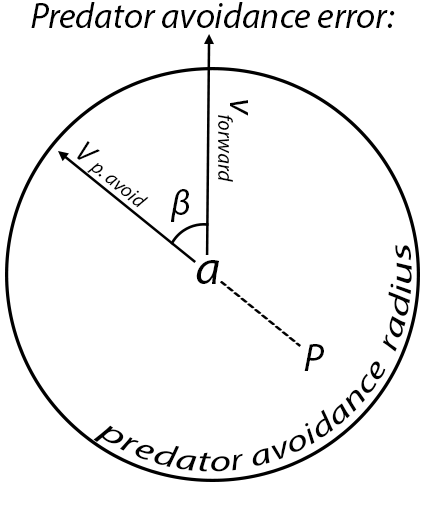}
    \includegraphics[scale=.3]{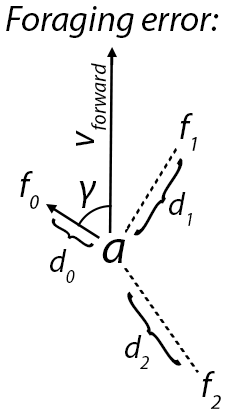}
    \caption{A micro level analysis of the five angular errors governing agent behavior.
}
    \label{fig:angleErrors}
\end{figure}

For each of the five rules, the angular errors are computed as follows (see figure \ref{fig:angleErrors}):\\

\emph{Angular error alignment:}
$ a_{0} $ is the agent of interest, $ a_{1} $ through $ a_{4} $ is the immediate neighborhood, \textit{Vforward} is the forward vector, \textit{Valign} is the control environment behavior vector, and $\phi$ is the alignment error angle.

\emph{Coherence angular error:}
$ a_{0} $ is the agent of interest, $ a_{1} $ to $ a_{5} $ are the neighbours, \textit{Vforward} is the forward vector, while \textit{Vcoh} is the vector of the control behavior and $\theta$ is the cohesion error angle.

\emph{Neighbor avoidance angular error:}
$ a_{0} $ is the agent of interest, $ a_{1} $ to $ a_{3} $ are the neighbours, \textit{Vn.avoid} represents the control neighbour avoidance vector, and $\lambda$ represents the neighbour avoidance error angle.

\emph{Predator avoidance angular error:}
The agent of interest is $ a $, the predator is \textit{P}, the forward vector is \textit{Vforward}, the ideal predator avoidance behavior vector is \textit{Vp.avoid}, and $\beta$ is the predator avoidance error angle.

\emph{Foraging angular error:}
$ a $ is the agent of interest, \textit{F0} to \textit{F2} are the food items, \textit{Vforward} is the forward vector, $ d_{0} $ to $ d_{2} $ are the distances to the food items, where $ d_{0} < d_{1} < d_{2} $ and $\gamma$ is the foraging error angle.
\\

In order to gain a realistic representation of the entire crowd, each collected metric was averaged across the total number of agents used in the simulation. The values were recorded 100 times, each recording after 100 consecutive frames, and then averaged over the total number of recordings. This is done in order to ensure that situations in which the agents exhibit behavioral artefacts are accounted for in the final measurements. The same procedure was applied to the predator data. Two populations were tested - one with fifteen agents and one with sixty agents - in order to determine whether the exhibited behavior changes linearly with the number of agents. Both ML models were trained for 1 million real-time steps. Accordingly, the length of the training sessions was determined by the relative time required for the algorithm to converge to an optimal solution, which is the point at which the agent's performance stops improving. A total of 4 food locations were selected, after trial and error, to ensure the greatest balance with regard to the other environmental elements. . Both models employed the same hyper-parameters in their neural networks.

\begin{figure}[h]
    \begin{center}
        \includegraphics[scale=.3]{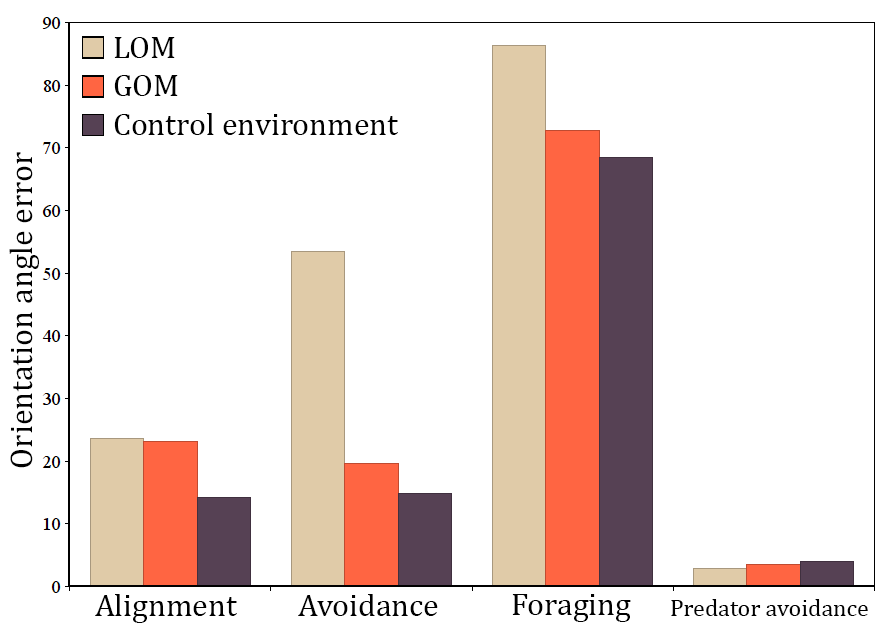}
        \includegraphics[scale=.3]{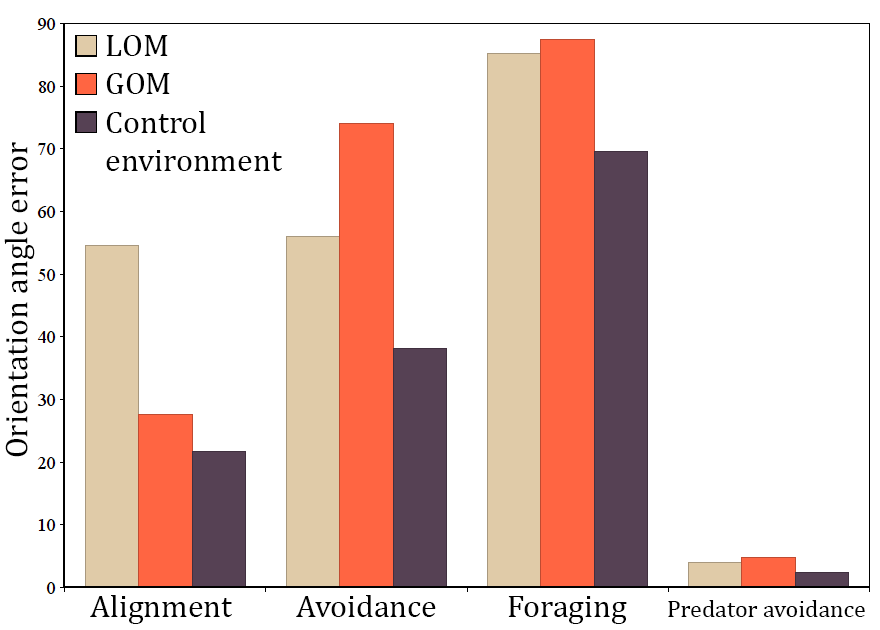}
        \caption{A comparison of alignment, avoidance, foraging, and predator avoidance errors with 15 agents per model (up) and 60 agents per model (down). Cohesion error values are less than 1 degree of difference, so they are considered insignificant.}
        \label{fig:angleErrors15and60}
    \end{center}
\end{figure}

\begin{figure}[h]
    \begin{center}
    
        \includegraphics[scale=.3]{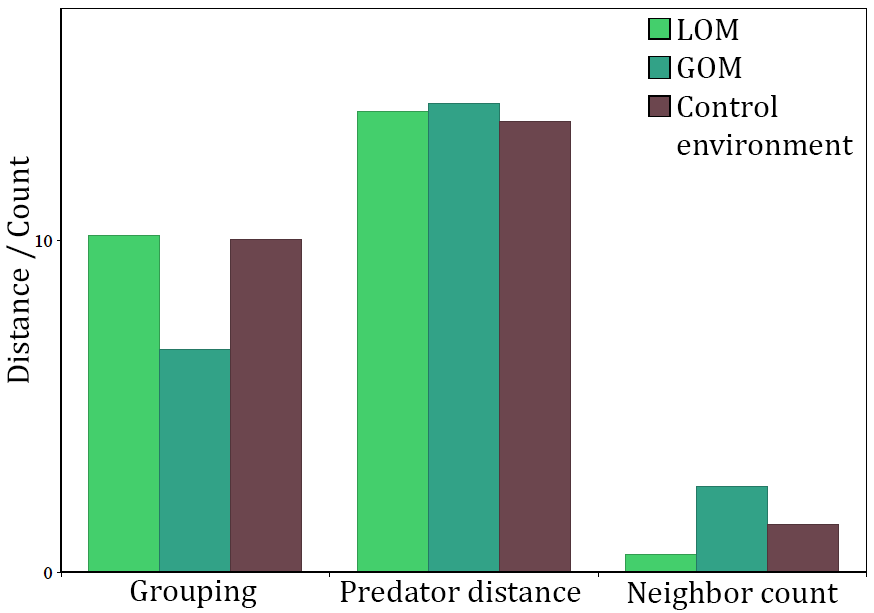} 
        \includegraphics[scale=.3]{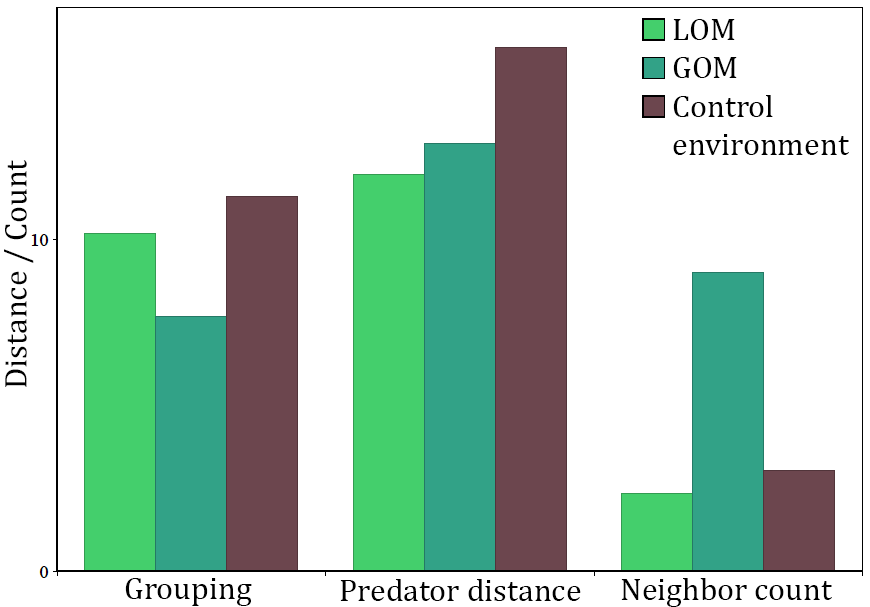} 
        \caption{A comparison of grouping (distance from nearest neighbour), predator distance and number of neighbors for 15 agents per model (up) and 60 agents per model (down). It is noteworthy that smaller distances indicate better performance for grouping, while low values indicate the exact opposite for the other metrics.}
        \label{fig:extraMetrics15and60}
    \end{center}
\end{figure}

In terms of alignment, avoidance, and foraging, the GOM portrays the poorer flocking behavior, followed by the LOM and the control environment (see figure \ref{fig:angleErrors15and60}).
The opposite is true in terms of predator avoidance and swarm cohesion. Using 60 agents, the GOM still outperforms the LOM in terms of alignment, but by a larger margin. The higher number of neighbours, however, reverses the trend related to neighbour avoidance and foraging, while retaining it related to predator avoidance. With 15 agents the LOM outperforms both the GOM and the control environment in grouping, predator distance and number of neighbours (see figure \ref{fig:extraMetrics15and60}). The GOM performs better than the control environment in terms of predator distance, but only marginal. A similar trend can be observed when using a higher number of agents.

\begin{figure}[h]
    \begin{center}
        \includegraphics[scale=.3]{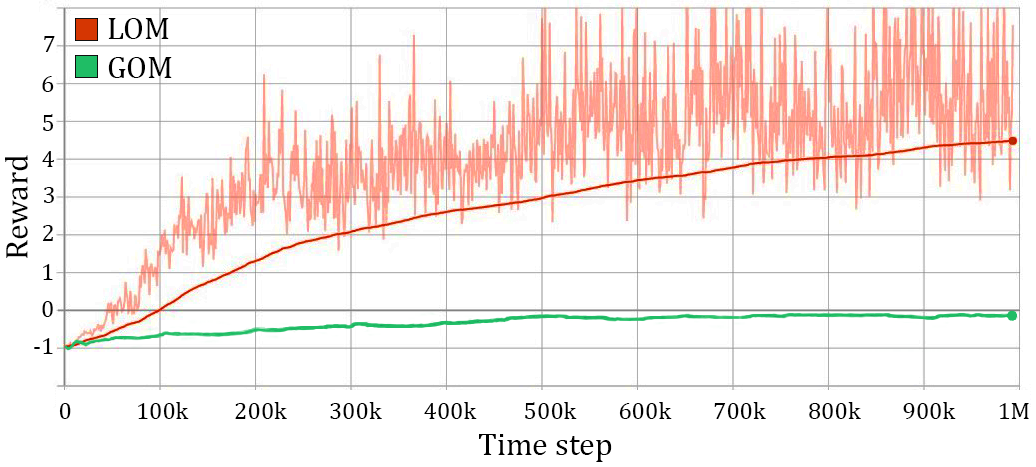} 
        \caption{The cumulative reward over the total number of training steps for the GOM (red) and LOM (green). Smoothing function applied due to high fluctuations.}
        \label{fig:cumulRewSmooth}
    \end{center}
\end{figure}

\begin{figure}[h]
    \begin{center}
        \includegraphics[scale=.3]{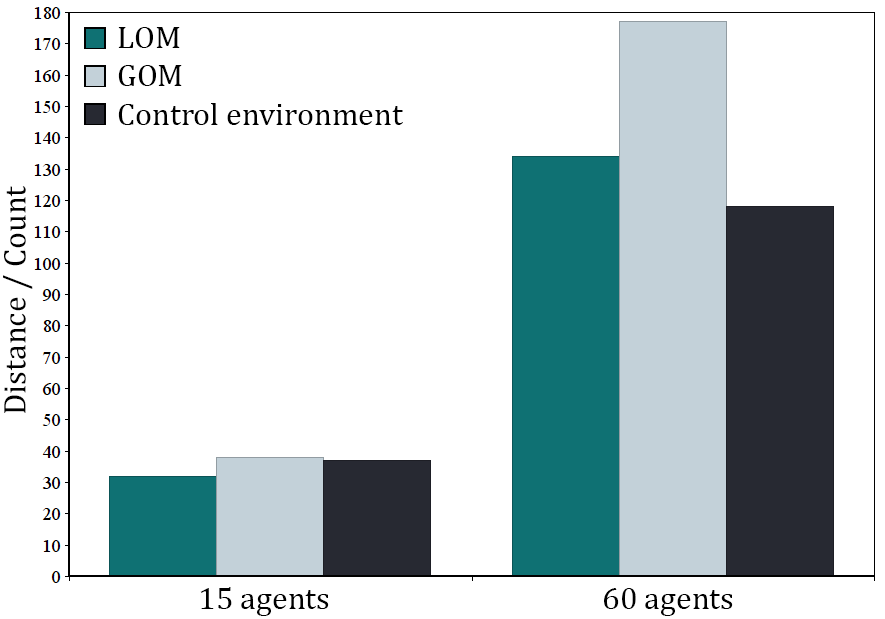} 
        \caption{Predator success rate over 10 000 steps}
        \label{fig:predTimesAte}
    \end{center}
\end{figure}

Throughout the training session, the GOM progresses continuously, stabilizing around 1.000.000 steps, maintaining positive reward the entire time (figure \ref{fig:cumulRewSmooth}). However, it generally does not result in the emergence of grouping behaviors. Even though the LOM displays near-perfect grouping behavior, it fails to obtain as high a reward, but still retains some progress over time. GOM exhibits the lowest number of casualties and is followed by the control environment (figure \ref{fig:predTimesAte}). Increasing the agent count by 60 results in an even greater difference.

\begin{figure}[h]
    \begin{center}
        \includegraphics[scale=.3]{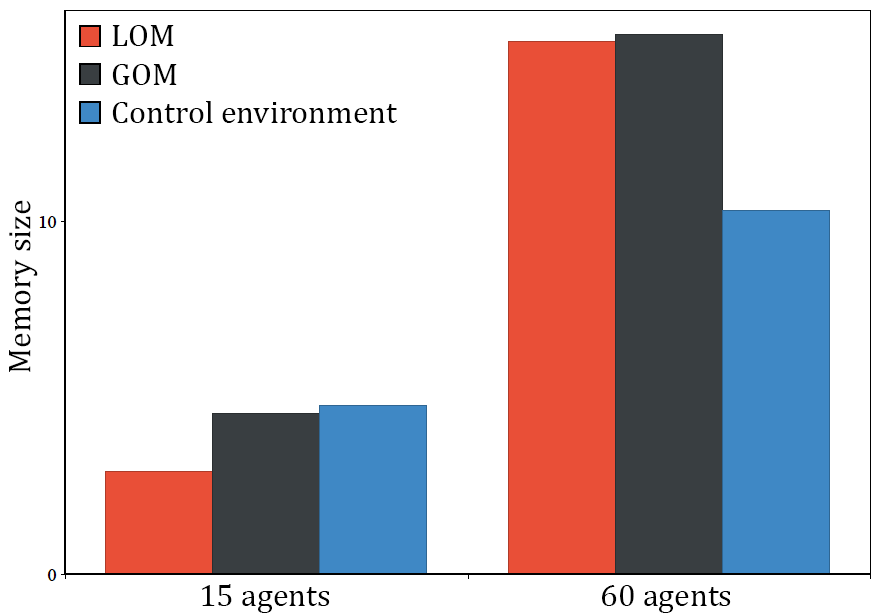} 
        \caption{Average predator memory size, over 10 000 steps}
        \label{fig:predMemory}
    \end{center}
\end{figure}

Memory size reflects confusion since it rises as more agents enter and flee the range of perception of the predator ( \ref{fig:predMemory}). As the number of agents increases, LOM and GOM predator confusion increases as well.

\begin{figure}[h]
    \begin{center}
        \includegraphics[scale=.3]{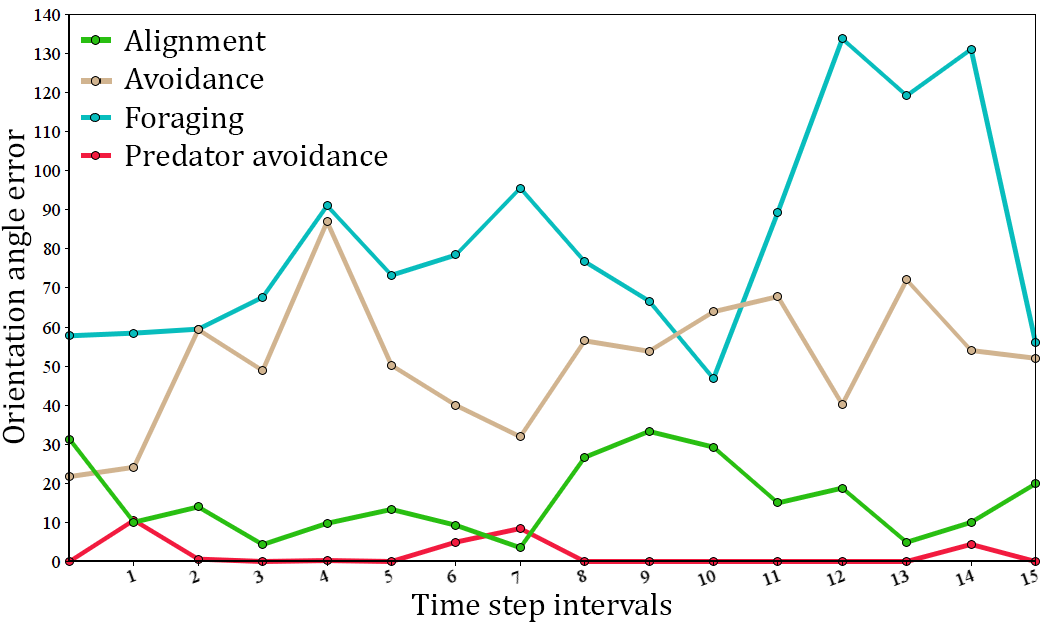} 
        \caption{Alignment, avoidance, foraging, and predator avoidance training progression. Each time step equals 50 000 training steps.}
        \label{fig:progression4Angles}
    \end{center}
\end{figure}

\begin{figure}[h]
    \begin{center}
        \includegraphics[scale=.3]{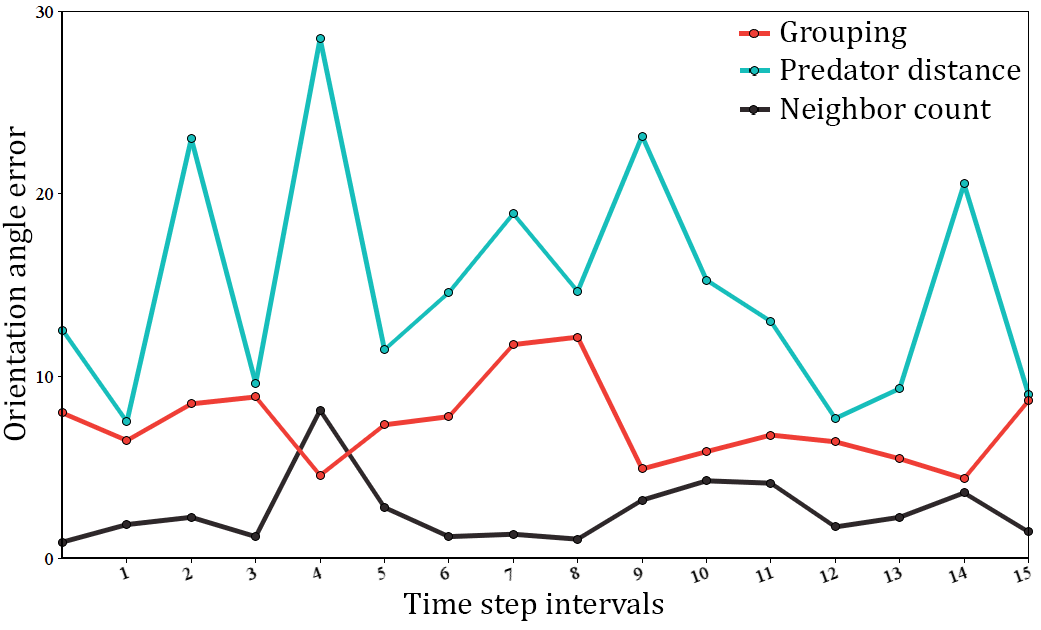} 
        \caption{GOM training progression in terms of grouping, predator avoidance and number of neighbors}
        \label{fig:progression4Angles1}
    \end{center}
\end{figure}

For neighbour avoidance and alignment, the GOM increases the angular error difference only once during the entire training (fig. \ref{fig:progression4Angles}. The predator avoidance is nearly perfect. As the cohesion never goes over 1 degree, it is ignored. While distance from the predator fluctuates greatly throughout the session, distance from the local group center and average neighbour number are relatively stable ( fig. \ref{fig:progression4Angles1}). The majority of agents are surrounded in a group almost always, with an average number of neighbors of 3.5.

\begin{figure*}
  \includegraphics[width=\textwidth,height=4cm]{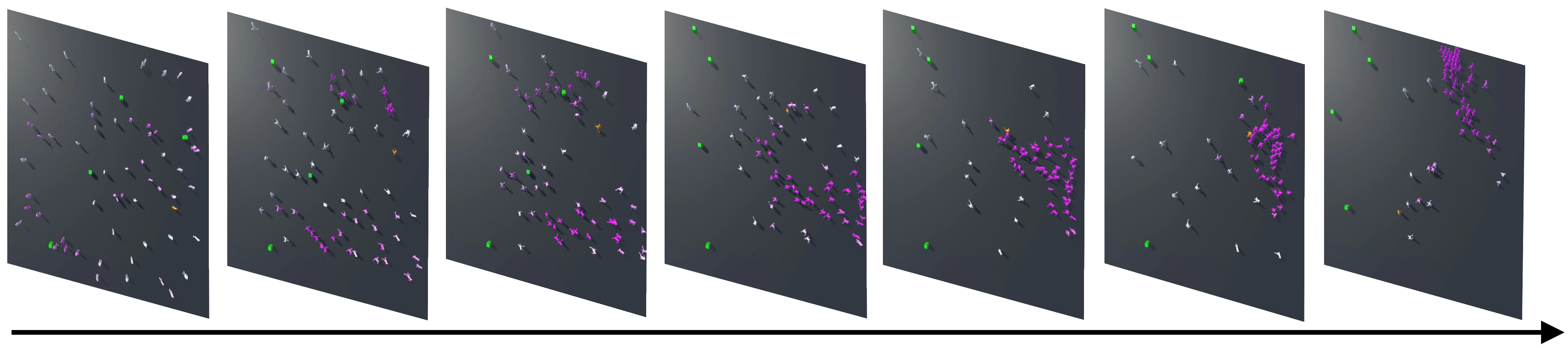}
  \caption{Grouping progression of the GOM over the first 60 time-steps. Agents are colored based on the size of their local groups (white to magenta). Upon initialization, flock members', food, and predator positions are randomized.}
\end{figure*}

\subsection{Collaborative Behavior}

LOM not only evolved towards a flocking behavior, but even surpassed the control environment in some of the measured criteria. The virtual creatures perform significantly poorer than the GOM in terms of searching for food, which fails to maximize the explicit reward. The algorithm would still prefer the 'safety in numbers' survival strategy despite the higher losses from the predator. The assumption is that this is a state of dynamic equilibrium between prolonging the life of the individual and the damage to the crowd as a whole the predator can inflict. The results demonstrate that complex emergent flocking behavior can be derived from a cooperative reinforcement learning environment by combining predatory threat, foraging need, and obstacle avoidance task. In addition, we illustrate how variations in information exchange lead to different global, collaborative behaviors. Longer training times are certain to show better results. It could also be argued that the system lacks the mesmerizing aesthetics of flock choreography that so fascinated Reynolds. Motion rigidity can be explained by the way information propagates between agents. The fact that biological agents process and act on the received information more slowly in a natural phenomenon can introduce information asymmetries, while propagation delays are negligible in an ideal simulated environment. 

\section{Conclusions}


A computational analysis was presented of the effects of predator-prey dynamics on group formation and its emergent social behavior. We show that the predator confusion hypothesis can play a role in evolving group formation. According to our results, dilution of risk can result from predator confusion. While the dilution of risk hypothesis underlies group formation, predator confusion effect is more important for the emergence of more complex social behaviors.

Agents that have global perception can perceive the whole world, while agents that have local perception can perceive only their immediate surroundings. Taking angular errors (\ref{fig:angleErrors15and60}) into account, we conclude that the GOM outperforms the LOM. As for their social behavior, the LOM agents adopted selfish, solitary survival strategies, while the GOM agents adopted a group survival strategy. As a result, the LOM was more robust and adaptive to population changes. But when it comes to their capacity to form groups, especially their distance from local group centers, distance from predators, and average number of neighbours, the GOM outperforms the LOM. Overall, the GOM produced better results than the control environment, as further evidenced by visual inspection.
\subsubsection{Group Formation}
Further, when comparing the average number of neighbors to the neighbor avoidance behavior, one can see that the GOM agents were so proficient at navigating in densely packed groups that they no longer needed to maintain the neighbour avoidance radius.  \ref{fig:extraMetrics15and60}. 


As a result of continuous collisions, the agents' velocity decreases, increasing their vulnerability. But physical bodies give the simulator a sense of realism usually lacking in similar works '. \cite{hahn2019emergent}, which usually treat flock members as pass-through objects, independent of collisions. Despite becoming slower when closer to other agents, the agents prioritized 'safety in numbers'.

With regards to predator distance, this strategy results in a better performance than the other ML model in this category \ref{fig:extraMetrics15and60}. By examining how many times the agents were caught by the predator, it is obvious that LOM performed better. It is clear from visual observations that despite never evolving into a group, the agents become more proficient at outmaneuvering the predator on their own.

GOM agents prioritize grouping to the point that once spawned, they seek out the nearest neighboring group in order to ensure their safety by joining the group. This suggests two major conclusions:  

\subsubsection{Dilution of Risk}
Dilution of risk factor appears sufficient to evolve group formation. 
The flock and predator also seem to have a well developed reaction-diffusion equilibrium, where change in activity immediately punished the inactive members of the flock. Dynamic improvement ensures that the algorithm does not plateau early on. Models instead encourage exploration resulting in complex, swarming survival strategies. 

\subsubsection{Global vs local perception}
Unlike LOM agents, the agents from GOM can see the predator at all times rather than relying on a perceptual sphere. Additionally, while GOM used independent behavior for the agents to balance foraging and predator avoidance, LOM prioritized predator avoidance and individual agents rarely broke away from the group. It is hypothesized that this is due to the fact that agents that are completely surrounded in a group have their ray cast vision obstructed by their immediate neighbours. In such a situation, the agent has two options: either follow the movement of its neighbours, thus maintaining a group formation, or break away from the group to better see the predator while simultaneously looking for food. As a result, the LOM agents prefer to avoid their neighbours and move as much as possible to increase their chances of finding food and decreasing their chances of missing incoming attacks. Despite obstructed vision, GOM agents are able to perceive the predator and can focus entirely on perfecting cohesion.

It appears that the higher predator memory size did not translate into catching fewer agents in most tests \ref{fig:cumulRewSmooth}. This is in contrast to the initial hypothesis that the safer the agents are, the more they are able to confuse the predator. Though less efficient in the long run, the agents still form the LOM-prioritized grouping. In other words, being in a group prolongs each individual agent's life, without reducing the predator's overall success. This can be reinforced by observing how the agents behave, since the most hunted-down flock members are frequently those who break away. Additionally, this is in line with predator satiation, as the 'safety in numbers' strategy primarily benefits the individual members of the group and not the group as a whole.

\footnotesize
\bibliographystyle{IEEEtran}
\bibliography{bibliography}

\begin{thebibliography}{10}
\providecommand{\url}[1]{#1}
\csname url@samestyle\endcsname
\providecommand{\newblock}{\relax}
\providecommand{\bibinfo}[2]{#2}
\providecommand{\BIBentrySTDinterwordspacing}{\spaceskip=0pt\relax}
\providecommand{\BIBentryALTinterwordstretchfactor}{4}
\providecommand{\BIBentryALTinterwordspacing}{\spaceskip=\fontdimen2\font plus
\BIBentryALTinterwordstretchfactor\fontdimen3\font minus
  \fontdimen4\font\relax}
\providecommand{\BIBforeignlanguage}[2]{{%
\expandafter\ifx\csname l@#1\endcsname\relax
\typeout{** WARNING: IEEEtran.bst: No hyphenation pattern has been}%
\typeout{** loaded for the language `#1'. Using the pattern for}%
\typeout{** the default language instead.}%
\else
\language=\csname l@#1\endcsname
\fi
#2}}
\providecommand{\BIBdecl}{\relax}
\BIBdecl

\bibitem{bedau2008emergence}
M.~A. Bedau and P.~E. Humphreys, \emph{Emergence: Contemporary readings in
  philosophy and science.}\hskip 1em plus 0.5em minus 0.4em\relax MIT press,
  2008.

\bibitem{cilliers1999complexity}
P.~Cilliers and D.~Spurrett, ``Complexity and post-modernism: Understanding
  complex systems,'' \emph{South African Journal of Philosophy}, vol.~18,
  no.~2, pp. 258--274, 1999.

\bibitem{ashby1968principles}
W.~R. Ashby, ``Principles of the self-organizing system,'' \emph{Modern systems
  research for the behavioral scientist}, pp. 108--118, 1968.

\bibitem{de2011self}
B.~De~Boer, ``Self-organization and language evolution,'' in \emph{The Oxford
  handbook of language evolution}.\hskip 1em plus 0.5em minus 0.4em\relax
  Oxford University Press, 2011.

\bibitem{turing1990chemical}
A.~M. Turing, ``The chemical basis of morphogenesis,'' \emph{Bulletin of
  mathematical biology}, vol.~52, no.~1, pp. 153--197, 1990.

\bibitem{camazine2003self}
S.~Camazine, J.-L. Deneubourg, N.~R. Franks, J.~Sneyd, E.~Bonabeau, and
  G.~Theraula, \emph{Self-organization in biological systems}.\hskip 1em plus
  0.5em minus 0.4em\relax Princeton university press, 2003, vol.~7.

\bibitem{kohonen1990self}
T.~Kohonen, ``The self-organizing map,'' \emph{Proceedings of the IEEE},
  vol.~78, no.~9, pp. 1464--1480, 1990.

\bibitem{maclennan1993synthetic}
B.~J. MacLennan and G.~M. Burghardt, ``Synthetic ethology and the evolution of
  cooperative communication,'' \emph{Adaptive behavior}, vol.~2, no.~2, pp.
  161--188, 1993.

\bibitem{zhou2010crowd}
S.~Zhou, D.~Chen, W.~Cai, L.~Luo, M.~Y.~H. Low, F.~Tian, V.~S.-H. Tay, D.~W.~S.
  Ong, and B.~D. Hamilton, ``Crowd modeling and simulation technologies,''
  \emph{ACM Transactions on Modeling and Computer Simulation (TOMACS)},
  vol.~20, no.~4, pp. 1--35, 2010.

\bibitem{reynolds1987flocks}
C.~W. Reynolds, ``Flocks, herds and schools: A distributed behavioral model,''
  in \emph{Proceedings of the 14th annual conference on Computer graphics and
  interactive techniques}, 1987, pp. 25--34.

\bibitem{reynolds1999steering}
------, ``Steering behaviors for autonomous characters,'' in \emph{Game
  developers conference}, vol. 1999.\hskip 1em plus 0.5em minus 0.4em\relax
  Citeseer, 1999, pp. 763--782.

\bibitem{dewi2011simulating}
M.~Dewi, M.~Hariadi, and M.~H. Purnomo, ``Simulating the movement of the crowd
  in an environment using flocking,'' in \emph{2011 2nd International
  Conference on Instrumentation, Communications, Information Technology, and
  Biomedical Engineering}.\hskip 1em plus 0.5em minus 0.4em\relax IEEE, 2011,
  pp. 186--191.

\bibitem{chiang2009emergent}
C.-S. Chiang, C.~Hoffmann, and S.~Mittal, ``Emergent crowd behavior,''
  \emph{Computer-Aided Design and Applications}, vol.~6, no.~6, pp. 865--875,
  2009.

\bibitem{kwasnicka2011flocking}
H.~Kwasnicka, U.~Markowska-Kaczmar, and M.~Mikosik, ``Flocking behaviour in
  simple ecosystems as a result of artificial evolution,'' \emph{Applied Soft
  Computing}, vol.~11, no.~1, pp. 982--990, 2011.

\bibitem{chang2019investigating}
G.~Chang and M.~Stjerndal, ``Investigating and modeling the emergent flocking
  behaviour of sheep under threat with fear contagion,'' 2019.

\bibitem{olson2013predator}
R.~S. Olson, A.~Hintze, F.~C. Dyer, D.~B. Knoester, and C.~Adami, ``Predator
  confusion is sufficient to evolve swarming behaviour,'' \emph{Journal of The
  Royal Society Interface}, vol.~10, no.~85, p. 20130305, 2013.

\bibitem{macaulay2016riot}
T.~Macaulay, \emph{RIoT control: understanding and managing risks and the
  internet of things}.\hskip 1em plus 0.5em minus 0.4em\relax Morgan Kaufmann,
  2016.

\bibitem{morrow1948schooling}
J.~E. Morrow~Jr, ``Schooling behavior in fishes,'' \emph{The Quarterly review
  of biology}, vol.~23, no.~1, pp. 27--38, 1948.

\bibitem{caraco1980avian}
T.~Caraco, S.~Martindale, and H.~R. Pulliam, ``Avian flocking in the presence
  of a predator,'' \emph{Nature}, vol. 285, no. 5764, pp. 400--401, 1980.

\bibitem{sullivan1981insect}
R.~T. Sullivan, ``Insect swarming and mating,'' \emph{The Florida
  Entomologist}, vol.~64, no.~1, pp. 44--65, 1981.

\bibitem{mamei2006case}
M.~Mamei, R.~Menezes, R.~Tolksdorf, and F.~Zambonelli, ``Case studies for
  self-organization in computer science,'' \emph{Journal of Systems
  Architecture}, vol.~52, no. 8-9, pp. 443--460, 2006.

\bibitem{lotka2002contribution}
A.~J. Lotka, ``Contribution to the theory of periodic reactions,'' \emph{The
  Journal of Physical Chemistry}, vol.~14, no.~3, pp. 271--274, 2002.

\bibitem{garvie2010spatiotemporal}
M.~R. Garvie and C.~Trenchea, ``Spatiotemporal dynamics of two generic
  predator--prey models,'' \emph{Journal of Biological dynamics}, vol.~4,
  no.~6, pp. 559--570, 2010.

\bibitem{winfree2001geometry}
A.~T. Winfree, \emph{The geometry of biological time}.\hskip 1em plus 0.5em
  minus 0.4em\relax Springer Science \& Business Media, 2001, vol.~12.

\bibitem{kopell1981target}
N.~Kopell and L.~Howard, ``Target pattern and spiral solutions to
  reaction-diffusion equations with more than one space dimension,''
  \emph{Advances in Applied Mathematics}, vol.~2, no.~4, pp. 417--449, 1981.

\bibitem{vanag2007localized}
V.~K. Vanag and I.~R. Epstein, ``Localized patterns in reaction-diffusion
  systems,'' \emph{Chaos: An Interdisciplinary Journal of Nonlinear Science},
  vol.~17, no.~3, p. 037110, 2007.

\bibitem{sherratt1997oscillations}
J.~A. Sherratt, B.~T. Eagan, and M.~A. Lewis, ``Oscillations and chaos behind
  predator--prey invasion: mathematical artifact or ecological reality?''
  \emph{Philosophical transactions of the Royal Society of London. Series B:
  Biological Sciences}, vol. 352, no. 1349, pp. 21--38, 1997.

\bibitem{witkowski2016emergence}
O.~Witkowski and T.~Ikegami, ``Emergence of swarming behavior: foraging agents
  evolve collective motion based on signaling,'' \emph{PloS one}, vol.~11,
  no.~4, p. e0152756, 2016.

\bibitem{haley2014exploring}
P.~Haley, R.~Olson, F.~Dyer, and C.~Adami, ``Exploring conditions that select
  for the evolution of cooperative group foraging,'' in \emph{ALIFE 14: The
  Fourteenth International Conference on the Synthesis and Simulation of Living
  Systems}.\hskip 1em plus 0.5em minus 0.4em\relax MIT Press, 2014, pp.
  310--311.

\bibitem{torney2011signalling}
C.~J. Torney, A.~Berdahl, and I.~D. Couzin, ``Signalling and the evolution of
  cooperative foraging in dynamic environments,'' \emph{PLoS computational
  biology}, vol.~7, no.~9, p. e1002194, 2011.

\bibitem{oboshi2003simulation}
T.~Oboshi, S.~Kato, A.~Mutoh, and H.~Itoh, ``A simulation study on the form of
  fish schooling for escape from predator,'' \emph{FORMA-TOKYO-}, vol.~18,
  no.~2, pp. 119--131, 2003.

\bibitem{demvsar2016balanced}
J.~Dem{\v{s}}ar, E.~{\v{S}}trumbelj, and I.~L. Bajec, ``A balanced mixture of
  antagonistic pressures promotes the evolution of parallel movement,''
  \emph{Scientific reports}, vol.~6, no.~1, pp. 1--12, 2016.

\bibitem{kunz2006prey}
H.~Kunz, T.~Z{\"u}blin, and C.~K. Hemelrijk, ``On prey grouping and predator
  confusion in artificial fish schools,'' in \emph{Proceedings of the Tenth
  International Conference of Artificial Life. MIT Press, Cambridge,
  Massachusetts}, 2006.

\bibitem{mills1982satiation}
N.~Mills, ``Satiation and the functional response: a test of a new model,''
  \emph{Ecological Entomology}, vol.~7, no.~3, pp. 305--315, 1982.

\bibitem{olson2016evolution}
R.~S. Olson, D.~B. Knoester, and C.~Adami, ``Evolution of swarming behavior is
  shaped by how predators attack,'' \emph{Artificial life}, vol.~22, no.~3, pp.
  299--318, 2016.

\bibitem{ruxton2008application}
G.~D. Ruxton and G.~Beauchamp, ``The application of genetic algorithms in
  behavioural ecology, illustrated with a model of anti-predator vigilance,''
  \emph{Journal of theoretical biology}, vol. 250, no.~3, pp. 435--448, 2008.

\bibitem{ficici1998challenges}
S.~G. Ficici and J.~B. Pollack, ``Challenges in coevolutionary learning:
  Arms-race dynamics,'' in \emph{Artificial Life VI: Proceedings of the sixth
  international conference on artificial life}, vol.~6.\hskip 1em plus 0.5em
  minus 0.4em\relax MIT Press, 1998, p. 238.

\bibitem{parunak1997go}
H.~V.~D. Parunak \emph{et~al.}, ``" go to the ant": Engineering principles from
  natural multi-agent systems,'' \emph{Ann. Oper. Res.}, vol.~75, pp. 69--101,
  1997.

\bibitem{schulman2017proximal}
J.~Schulman, F.~Wolski, P.~Dhariwal, A.~Radford, and O.~Klimov, ``Proximal
  policy optimization algorithms,'' \emph{arXiv preprint arXiv:1707.06347},
  2017.

\bibitem{hahn2019emergent}
C.~Hahn, T.~Phan, T.~Gabor, L.~Belzner, and C.~Linnhoff-Popien, ``Emergent
  escape-based flocking behavior using multi-agent reinforcement learning,'' in
  \emph{The 2018 Conference on Artificial Life: A Hybrid of the European
  Conference on Artificial Life (ECAL) and the International Conference on the
  Synthesis and Simulation of Living Systems (ALIFE)}.\hskip 1em plus 0.5em
  minus 0.4em\relax MIT Press, 2019, pp. 598--605.

\bibitem{hamburger1973n}
H.~Hamburger, ``N-person prisoner's dilemma,'' \emph{Journal of Mathematical
  Sociology}, vol.~3, no.~1, pp. 27--48, 1973.

\bibitem{rossi2018review}
F.~Rossi, S.~Bandyopadhyay, M.~Wolf, and M.~Pavone, ``Review of multi-agent
  algorithms for collective behavior: a structural taxonomy,''
  \emph{IFAC-PapersOnLine}, vol.~51, no.~12, pp. 112--117, 2018.

\bibitem{back1993overview}
T.~B{\"a}ck and H.-P. Schwefel, ``An overview of evolutionary algorithms for
  parameter optimization,'' \emph{Evolutionary computation}, vol.~1, no.~1, pp.
  1--23, 1993.

\bibitem{coggan2004exploration}
M.~Coggan, ``Exploration and exploitation in reinforcement learning,''
  \emph{Research supervised by Prof. Doina Precup, CRA-W DMP Project at McGill
  University}, 2004.

\bibitem{sutton2018reinforcement}
R.~S. Sutton and A.~G. Barto, \emph{Reinforcement learning: An
  introduction}.\hskip 1em plus 0.5em minus 0.4em\relax MIT press, 2018.

\bibitem{juliani2018unity}
A.~Juliani, V.-P. Berges, E.~Vckay, Y.~Gao, H.~Henry, M.~Mattar, and D.~Lange,
  ``Unity: A general platform for intelligent agents,'' \emph{arXiv preprint
  arXiv:1809.02627}, 2018.

\end{thebibliography}

\end{document}